\documentclass[runningheads,a4paper]{llncs}
\usepackage{amssymb}
\usepackage{amsmath}
\usepackage{array}
\usepackage[hidelinks]{hyperref}
\usepackage{graphicx}
\usepackage{tikz}

\makeatletter

\def\define#1{\@ifnextchar [{\@MYargdef#1}{\@MYargdef#1[0]}}

\def\@MYargdef#1[#2]#3{\@ifdefinable #1{\@MYreargdef#1[#2]{#3}}}

\def\redefine#1{\edef\@tempa{\expandafter\@cdr\string 
  #1\@nil}\@ifundefined{\@tempa}{\@latexerr{\string#1\space undefined}\@ehc
    }{}\@ifnextchar [{\@MYreargdef#1}{\@MYreargdef#1[0]}}

\catcode`\?=11\relax
\def\@MYreargdef#1[#2]#3{\@tempcnta#2\relax\let#1\relax
\edef\@tempa{\def#1}\@tempcntb \@ne
\let\@?@?\relax\@whilenum\@tempcnta>0
\do{\edef\@tempa{\@tempa\@?@?\the\@tempcntb}\advance\@tempcntb \@ne \advance
\@tempcnta \m@ne}\let\@?@?##\@tempa{#3}}
\catcode`\?=12\relax

\makeatother

\renewcommand{\bmod}%
{\mskip-\medmuskip \mkern5mu \mathbin{\idFont mod} \mkern5mu \mskip-\medmuskip}

\define{\mathsym}[1]{\relax\ifmmode#1\else
	\errmessage{Mathematical symbol outside math mode}\fi}
\define{\idFont}{\sf}
\define{\id}[1]{\mathsym{{\idFont #1}}}

\define{\m}[1]{$#1$}
\define{\M}[1]{$$#1$$}

\define{\gt}{\mathsym{\mathchar 12606\relax}} 
\define{\lt}{\mathsym{\mathchar 12604\relax}} 

\define{\paren}[1]{(#1)}
\define{\parenA}[1]{(#1)}
\define{\parenB}[1]{\bigl(#1\bigr)}
\define{\parenC}[1]{\Bigl(#1\Bigr)}
\define{\parenAuto}[1]{\left(#1\right)}
\define{\set}[1]{\lbrace#1\rbrace}
\define{\setA}[1]{\lbrace#1\rbrace}
\define{\setB}[1]{\bigl\lbrace#1\bigr\rbrace}
\define{\setC}[1]{\Bigl\lbrace#1\Bigr\rbrace}
\define{\setCond}[2]{\lbrace#1\mathrel{\vert}#2\rbrace}
\define{\setCondA}[2]{\lbrace#1\mathrel{\vert}#2\rbrace}
\define{\setCondB}[2]{\bigl\lbrace#1\bigm\vert#2\bigr\rbrace}
\define{\setCondC}[2]{\Bigl\lbrace#1\Bigm\vert#2\Bigr\rbrace}

\define{\union}{\cup}
\define{\unionUntil}{\union\cdots\union}
\define{\unionMulti}[2]{\bigcup_{#1}^{#2}}
\define{\intersect}{\cap}
\define{\intersectUntil}{\intersect\cdots\intersect}
\define{\intersectMulti}[2]{\bigcap_{#1}^{#2}}
\define{\timesUntil}{\times\cdots\times}
\define{\setsize}[1]{{\left\vert #1 \right\vert}}
\define{\compl}[1]{\overline{#1}}
\renewcommand{\setminus}{-}
\define{\powerset}[1]{2^{#1}}
\define{\inB}{\;\in\;}

\define{\defEq}{:=}
\define{\defEqB}{\;:=\;}
\define{\defIff}{\;\mathsym{:\Longleftrightarrow}\;}
\define{\metaThen}{\mathsym{\;\Longrightarrow\;}}
\define{\metaIf}{\mathsym{\;\Longleftarrow\;}}
\define{\metaIff}{\mathsym{\;\Longleftrightarrow\;}}

\define{\lneg}{\mathsym{\neg}}
\define{\lif}{\mathsym{\leftarrow}}
\define{\lifDup}{\mathsym{\Leftarrow}}
\define{\lthen}{\mathsym{\rightarrow}}
\define{\liff}{\mathsym{\leftrightarrow}}
\define{\lfalse}{\id{false}}
\define{\ltrue}{\id{true}}

\define{\lorUntil}{\lor\cdots\lor}
\define{\landUntil}{\land\cdots\land}
\define{\lorMulti}[2]{\bigvee_{#1}^{#2}}
\define{\landMulti}[2]{\bigwedge_{#1}^{#2}}

\define{\lfor}[4]{#1\:#2\:#3:#4}
\define{\lexistsQ}{\exists}
\define{\lexists}[3]{\lfor{\lexistsQ}{#2}{#1}{#3}}
\define{\lallQ}{\forall}
\define{\lall}[3]{\lfor{\lallQ}{#2}{#1}{#3}}

\define{\answerPred}{\id{answer}}

\define{\natNum}{\mathsym{{\rm l\kern-0.13em N}}}

\define{\realNum}{\mathsym{{\rm I\kern-0.14em R}}}

\define{\struct}[1]{\langle#1\rangle}
\define{\twoCases}[4]{\left\{\begin{array}{l@{\kern10pt}l}
				#1&\mbox{#2}\\#3&\mbox{#4}
				\end{array}\right.}
\define{\until}{, \ldots,}
\define{\seqOf}[1]{#1^*}
\define{\emptySeq}{\epsilon}
\define{\w}{\id{w}}

\define{\code}[1]{{\tt #1}} 

{\begin{center}\begin{tt}\begin{tabular}{@{}l@{}}}%
{\end{tabular}\end{tt}\end{center}}

	{\begin{center}\begin{tt}\begin{tabular}{@{}l@{ }l@{}}}%
	{\end{tabular}\end{tt}\end{center}}

\define{\U}{{\char95}} 
\define{\LT}{{\char60}} 
\define{\GT}{{\char62}} 
\define{\B}{{\char92}} 
\define{\AMP}{{\char38}} 
\define{\D}{{\char36}} 
\define{\SN}{{\char126}} 
\define{\HASH}{{\char35}} 
\define{\Q}{{\char34}} 
\define{\PCT}{{\char37}} 
\define{\LB}{{\char123}} 
\define{\RB}{{\char125}} 
\define{\HAT}{{\char94}} 

\define{\ALPH}{\id{ALP\kern-0.08em H}}
\define{\LOG}{\id{LOG}}
\define{\VARS}{\id{V\kern-0.17em ARS}}
\define{\var}{\mathsym{v}}
\define{\varA}{\id{X}}
\define{\varB}{\id{Y}}
\define{\varC}{\id{Z}}
\define{\Vars}{\mathsym{{\cal V}}}
\define{\const}{\mathsym{c}}
\define{\constA}{\id{a}}
\define{\constB}{\id{b}}
\define{\constC}{\id{c}}
\define{\data}{\mathsym{d}}
\define{\dataSet}{\mathsym{{\cal D}}}
\define{\sig}{\Sigma}
\define{\SORTS}{\id{{\cal S}}}
\define{\sort}{\id{s}}
\define{\PREDS}{\id{{\cal P}}}
\define{\pred}{\id{p}}
\define{\predA}{\id{p}}
\define{\predB}{\id{q}}
\define{\predC}{\id{r}}
\define{\arity}{\id{n}}
\define{\level}{l}
\define{\FUNS}{\id{{\cal F}}}
\define{\fun}{\id{f}}
\define{\args}{\alpha}
\define{\argsB}{\beta}
\define{\argSorts}{\alpha}
\define{\resSort}{\rho}
\define{\interp}{\id{{\cal I}}}
\define{\interpB}{\id{{\cal J}}}
\define{\ass}{\id{{\cal A}}}
\define{\iV}[2]{(#1,#2)}
\define{\eval}[2]{#1\lbrack\kern-0.15em\lbrack#2\rbrack\kern-0.15em\rbrack}
\define{\evalV}[3]{\eval{\iV{#1}{#2}}{#3}}
\define{\varDecl}{\nu}
\define{\TERMS}{\id{T\kern-0.1em E}}
\define{\term}{\id{t}}
\define{\termB}{\id{u}}
\define{\argA}{\id{a}}
\define{\argB}{\id{b}}
\define{\argC}{\id{c}}
\define{\AT}{\id{AT}}
\define{\FO}{\id{FO}}
\define{\fo}{\varphi}
\define{\foB}{\psi}
\define{\fos}{\Phi}
\define{\modify}[3]{#1\langle#2/#3\rangle}
\define{\impl}{\vdash}
\define{\subst}{\theta}
\define{\substB}{\sigma}
\define{\mgu}{\id{mgu}}
\define{\doSubst}[2]{#2\,#1}
\define{\doSubstB}[2]{(#2)\,#1}
\define{\HU}{\id{{\cal U}}}
\define{\HB}{\id{{\cal B}}}
\define{\HSet}{\id{H}}
\define{\fact}{\id{F}}
\define{\ground}{\id{ground}}
\define{\LAT}{\id{{\cal M}}}
\define{\LATB}{\id{{\cal N}}}

\define{\emptyClause}{{\hbox{%
	\setlength{\unitlength}{0.24ex}%
	\begin{picture}(5,5)(0,0)
	\put(0,0){\line(0,1){5}}
	\put(0,0){\line(1,0){5}}
	\put(0,5){\line(1,0){5}}
	\put(5,0){\line(0,1){5}}
	\end{picture}}}}

\define{\lit}{\id{L}}
\define{\litA}{\id{A}}
\define{\litB}{\id{B}}
\define{\litC}{\id{C}}
\define{\Body}{\mathsym{{\cal B}}}
\define{\F}{\id{F}} 
\define{\ruleFun}[1]{\id{r}_{#1}}

\define{\prog}{\mathsym{{\idFont P}}}
\define{\T}[1]{\mathsym{{\idFont T}}_{#1}}
\define{\TP}{\T{\prog}}
\define{\Tneg}[2]{\mathsym{{\idFont T}}_{#1,#2}}
\define{\lub}{\id{lub}}
\define{\glb}{\id{glb}}
\define{\lfp}{\id{l\kern-0.1em f\kern-0.1em p}}

\define{\I}{\id{{\cal I}}}
\define{\J}{\id{{\cal J}}}

\define{\nf}{\mathop{\id{not\kern0.2em}}}
\define{\nfPred}[1]{\mathop{\id{not{\U}}#1}}

\define{\free}{\id{f}}
\define{\bound}{\id{b}}

\define{\bp}{\id{bp}}
\define{\binding}{\beta}
\define{\bindingSet}{{\cal B}}

\define{\vars}{\mathsym{{\cal X}}}
\define{\varsA}{\mathsym{{\cal X}}}
\define{\varsB}{\mathsym{{\cal Y}}}
\define{\varsC}{\mathsym{{\cal Z}}}

\define{\inputVars}{\id{input}}

\define{\freeVar}{\id{vars}}
\define{\varsOf}{\id{vars}}
\define{\boundPos}{\id{bound}^+}
\define{\boundNeg}{\id{bound}^-}
\define{\unboundPos}{\id{unbound}^+}
\define{\unboundNeg}{\id{unbound}^-}
\define{\err}{\id{err}}

\newcounter{ProgramLine}
\newcounter{FirstLine}

\newenvironment{progTabular}{%
	\addtocounter{ProgramLine}{-1}%
	\setcounter{FirstLine}{1}%
	\ifvmode\vspace{5mm}\fi%
	\begin{list}{\(\bullet\)\hfill}{
		\parskip 3pt plus 1pt
		\labelwidth 0pt
		\labelsep 0pt
		\leftmargin 8mm
		\listparindent \parindent
		\topsep 4mm
		\parsep 3pt plus 1pt
		\itemsep 0pt
		\partopsep 0pt
		\itemindent 0pt
		\rightmargin 0pt
	}
	\item[]
	\begin{tabular}{@{}r@{\hspace{2mm}}l@{}}}%
	{\end{tabular}%
		\end{list}}

\newenvironment{progPart}[1]{%
		\setcounter{ProgramLine}{#1}%
		\begin{progTabular}}%
	{\end{progTabular}}

	{\end{progTabular}%
		\end{tt}}

	{\end{progTabular}%
		\end{tt}}

\define{\tabX}[1]{\ifnum\value{FirstLine}=1\setcounter{FirstLine}{0}\else
	\ifhmode\\[0.5pt]\fi\fi
	\stepcounter{ProgramLine}(\arabic{ProgramLine})&
	\hspace*{#1}}
\define{\tabA}{\tabX{0cm}}
\define{\tabB}{\tabX{8mm}}
\define{\tabC}{\tabX{16mm}}
\define{\tabD}{\tabX{24mm}}
\define{\tabE}{\tabX{32mm}}
\define{\tabF}{\tabX{40mm}}
\define{\tabG}{\tabX{50mm}}
\define{\tabH}{\tabX{60mm}}
\define{\tabBox}[1]{\vrule height0pt depth0pt width0pt\hbox to14mm{#1\ \hfil}}

\define{\IF}{{\bf if }}
\define{\THEN}{{\bf then }}
\define{\ELSE}{{\bf else }}
\define{\FI}{{\bf fi}}
\define{\FOREACH}{{\bf foreach }}
\define{\FOR}{{\bf for }}
\define{\TO}{{\bf to }}
\define{\FROM}{{\bf from }}
\define{\IN}{{\bf in }}
\define{\WITH}{{\bf with }}
\define{\AND}{{\bf and }}
\define{\NOT}{{\bf not }}
\define{\TRUE}{{\bf true}}
\define{\FALSE}{{\bf false}}
\define{\DO}{{\bf do }}
\define{\OD}{{\bf od}}
\define{\WHILE}{{\bf while }}
\define{\BREAK}{{\bf break}}
\define{\PROCEDURE}{{\bf procedure }}
\define{\BEGIN}{{\bf \code{\LB}} }
\define{\END}{{\bf \code{\RB}} }
\define{\RETURN}{{\bf return }}
\define{\LET}{{\bf let }}
\define{\NEW}{{\bf new }}
\define{\COMPUTE}{{\bf compute }}
\define{\APPEND}{{\bf append }}
\define{\PRINT}{{\bf print }}
\define{\BOOL}{{\bf bool }}
\define{\NIL}{{\bf nil }}
\define{\OUTPUT}{{\bf output }}
\define{\INSERT}{{\bf insert }}
\define{\INTO}{{\bf into }}
\define{\CHOOSE}{{\bf choose }}

\define{\COMMENT}[1]{{\rm /\raisebox{-.6ex}{*} #1 \raisebox{-.6ex}{*}/}}

\define{\CPP}{C{\tt ++}}

\define{\activePart}[1]{\id{a}(#1)}
\define{\delayedPart}[1]{\id{d}(#1)}

\define{\lineno}[1]{\hbox to 1.6em{\hfil\m{\lbrack#1\rbrack}}}
\define{\lineref}[1]{\m{\lbrack#1\rbrack}}
\define{\comment}[1]{\mbox{// #1}}

\define{\C}{\mathsym{C}}

\define{\edbPredA}{\id{e}}
\define{\edbPredB}{\id{r}}
\define{\edbPredC}{\id{s}}

\define{\state}{\mathsym{{\cal S}}}
\define{\Rule}{\mathsym{{R}}}
\define{\db}{\mathsym{{\cal D}}}

\define{\vertices}{\mathsym{{\cal V}}}
\define{\edges}{\mathsym{{\cal E}}}

\define{\called}[2]{(\lineref{#1},\lineref{#2})}

\define{\edge}{\id{edge}}
\define{\mirrorNode}{\id{mirror\_node}}
\define{\nodeA}{\id{a}}
\define{\nodeB}{\id{b}}

\define{\grandparent}{\id{grandparent}}
\define{\parent}{\id{parent}}
\define{\father}{\id{father}}
\define{\mother}{\id{mother}}
\define{\personA}{\id{ann}}
\define{\personB}{\id{betty}}
\define{\personC}{\id{chris}}
\define{\personD}{\id{david}}

\define{\instRel}{\mathsym{\rightarrow_I}}
\define{\redRel}{\mathsym{\rightarrow_R}}
\define{\redRelDB}[1]{\mathsym{\rightarrow_{R(#1)}}}
\define{\anyRel}{\mathsym{\rightarrow}}
\define{\instClosure}{\id{Inst}^+}
\define{\redClosure}[1]{\id{Red}^+_{#1}}
\define{\transRel}{\mathsym{\rightarrow_T}}

\define{\query}{\id{Q}}
\define{\varS}{\id{V}}
\define{\std}{\id{std}}
\define{\sel}{\id{sel}}
\define{\call}{\id{call}}
\define{\csR}{\id{R}}
\define{\csD}{\id{D}}
\define{\predD}{\id{s}}
\define{\relU}{\mapsto_U} 
\define{\relD}{\mapsto_D} 
\define{\relC}{\mapsto_C} 
\define{\relP}{\mapsto_P} 
\define{\relN}{\mapsto_N} 
\define{\relS}{\mapsto_S} 
\define{\relF}{\mapsto_F} 
\define{\relR}{\mapsto_R} 
\define{\relE}{\mapsto_E} 

\define{\AAA}{\mathsym{{\cal A}}}
\define{\BBB}{\mathsym{{\cal B}}}
\define{\CCC}{\mathsym{{\cal C}}}

\define{\ArgPos}{\mathsym{{\cal A}}}
\define{\Types}{\mathsym{{\cal T}}}
\define{\type}{\mathsym{\tau}}
\define{\typeSpec}{\mathsym{{\cal S}}}
\define{\edbTypeSpec}{\mathsym{{\cal S}_E}}
\define{\typeIntersect}{\mathsym{\sqcap}}
\define{\typeBot}{\mathsym{\bot}}
\define{\subtype}{\mathsym{\sqsubseteq}}
\define{\typeAss}{\mathsym{\theta}}
\define{\typeT}[1]{\mathsym{D}}

\begin{document}

\title{A Rule-Based Approach to Analyzing Database Schema Objects with Datalog}
\titlerunning{Analyzing Database Schema Objects with Datalog}
 \author{Christiane Engels\inst{1}, Andreas Behrend\inst{1} \and Stefan Brass\inst{2}}
 \institute{Rheinische Friedrich-Wilhelms-Universit\"at Bonn,
        Institut f\"ur Informatik III,\\
 	R\"omerstra\ss e 164, D-53117 Bonn, Germany\\
 	\email{\{engelsc, behrend\} @ cs.uni-bonn.de}
	\and
	Martin-Luther-Universit\"at Halle-Wittenberg,
	Institut f\"ur Informatik,\\
	Von-Seckendorff-Platz~1, D-06099 Halle, Germany\\
	\email{brass@informatik.uni-halle.de}}




\maketitle


\begin{abstract}
Database schema elements such as tables, views, triggers and functions are typically defined with many interrelationships.
In order to support database users in understanding a given schema, a rule-based approach for analyzing the respective dependencies is proposed using Datalog expressions.
We show that many interesting properties of schema elements can be systematically determined this way.
The expressiveness of the proposed analysis is exemplarily shown with the problem of computing induced functional dependencies for derived relations.
The propagation of functional dependencies plays an important role in data integration and query optimization but represents an undecidable problem in general.
And yet, our rule-based analysis covers all relational operators as well as linear recursive expressions in a systematic way showing the depth of analysis possible by our proposal. The analysis of functional dependencies is well-integrated in a uniform approach to analyzing dependencies between schema elements in general.
\end{abstract}


\section{Introduction}
\label{Introduction}
%
The analysis of database schema elements such as tables, views, triggers, user-defined functions and constraints provides valuable information for database users for understanding, maintaining and managing a database application and its evolution.
In the literature, schema analysis has been investigated for improving the quality of SQL/program code or detecting program errors~\cite{BG05}, for detecting the consequences of schema changes~\cite{MER08}, for versioning~\cite{HVBRL17}, and matching~\cite{MZ98}.
In addition, the analysis of schema objects plays an important role for tuning resp. refactoring database applications~\cite{BSSW07}. 
All these approaches rely on exploring dependencies between schema objects and an in-depth analysis of their components and interactions.
A comprehensive and flexible analysis of schema elements, however, is not provided as these approaches are typically restricted to some subparts of a given schema.
The same is true for analysis features provided by commercial systems where approaches such as integrity checking, executing referential actions or query change notification (as provided by Oracle) already use schema object dependencies but in an implicit and nontransparent way, only.
That is, no access to the underlying meta-data is provided to the user nor can be freely analyzed by means of user-defined queries.
Even the meta-data about tables and SQL views which are sometimes provided by system tables cover only certain information of the respective schema elements.
This makes it difficult for database users to understand a given schema, explain specific derivations or oversee the consequences of intended schema modifications. 

In this paper, we propose a uniform approach for analyzing schema elements in a comprehensive way. To this end, the schema objects are compiled and their meta-data is stored into a Datalog program which employs queries for deriving interesting properties of the schema. 
This way, indirect dependencies between tables, views and user-defined functions can be determined which is important for understanding follow-up changes. In order to show the expressiveness of the proposed analysis, our rule-based approach is applied to the 
problem of deducing functional dependencies (FDs) for derived relations, i.e., views, based on FDs defined for base relations. This so-called \textit{FD propagation} or \textit{FD-FD implication} problem has been studied since the 80s~\cite{BGHP98,Fan08,Klug80,KlPr82,PBPP03,WaYu92}
and has applications in data exchange~\cite{FKPT09}, data integration~\cite{CCGL04}, data cleaning~\cite{Fan08}, data transformations~\cite{DFHQ03}, and semantic query optimization~\cite{PBPP02}.

Functional dependencies describe relationships between attributes of a data\-base relation and are the most widely used uni-relational dependencies~\cite{DeAd85}. 
They arise naturally in many ways, for instance when modeling key constraints, one-to-one or one-to-many relationships.
The problem of FD propagation is undecidable in the general setting and coNP-complete for many special cases~\cite{Fan08}.
Consequently, the task of finding induced FDs is rather complex and needs to be flexible in order to allow for further refinements. 
We show that our rule-based approach to schema analysis is well-suited for realizing techniques for FD propagation in a declarative way indicating the expressiveness of the proposed analysis.
In particular, our contributions are as follows:
\begin{itemize}
\item We propose an approach for analyzing the properties of views, tables, triggers and functions in a uniform way.
\item Our declarative approach can be easily extended for refining the analysis by user-defined queries.
\item The employed Datalog solution can be simply transfered into SQL systems.
\item In order to show the expressiveness of our approach, the implication problem for functional dependencies is investigated using our approach.
\end{itemize}
The paper is organized as follows: First, we introduce the rule-based framework for analyzing schema objects in Section~\ref{Schema Analysis}.
Afterwards, the problem of FD propagation is investigated serving as a use case in Section~\ref{Functional Dependency Propagation}.
In this section, a systematic way for deriving FD propagation rules is developed (Subsection~\ref{Propagation Rules}) before the most difficult operations 'union' and 'recursion' are investigated in more detail in Subsection~\ref{Union and Recursion}. 
Finally, we draw a conclusion in Section \ref{Conclusion}.

\section{Rule-Based Schema Analysis}
\label{Schema Analysis}
%
\begin{figure}[t]
\center
\includegraphics[width=6cm]{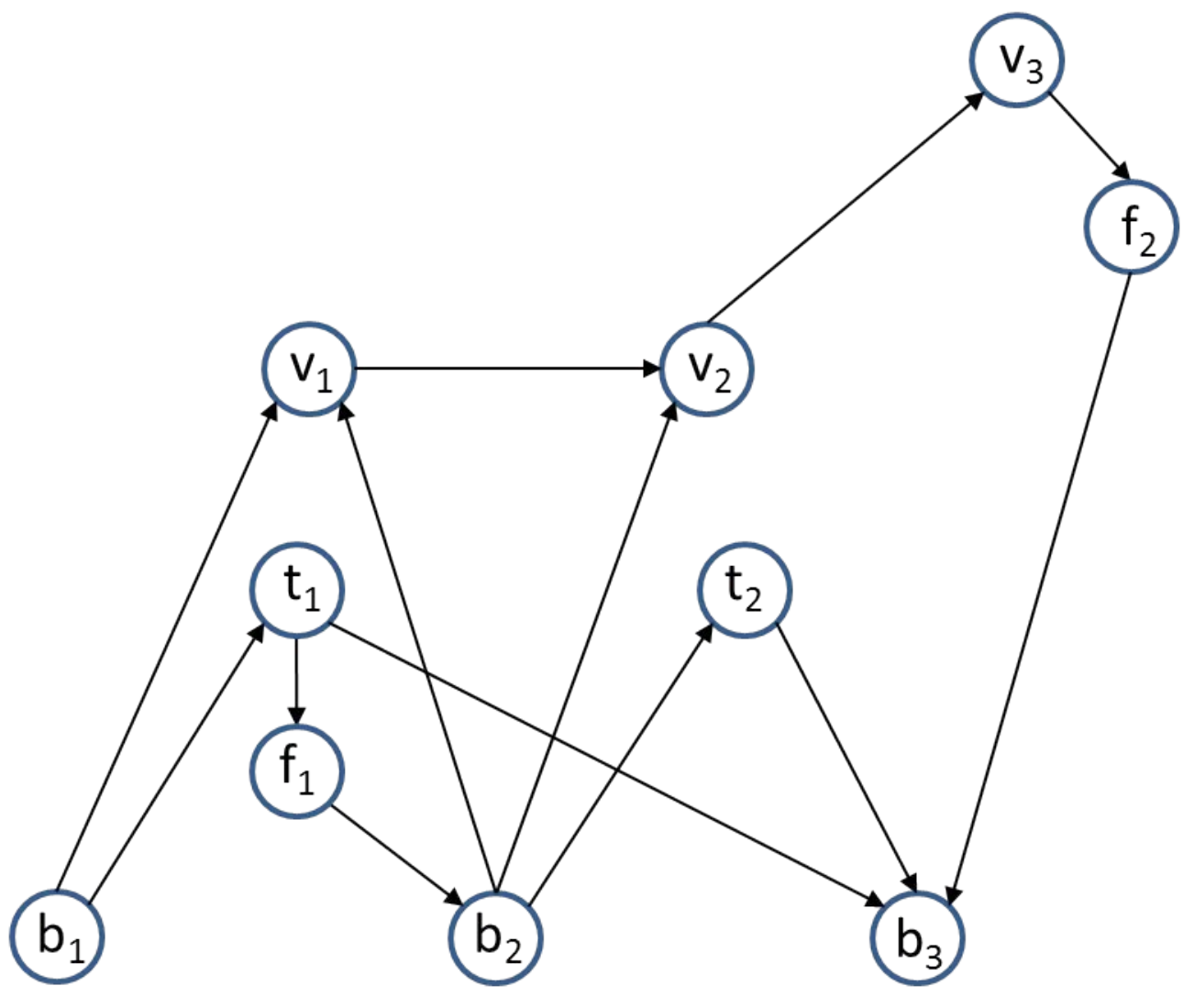}%
\caption{Dependency graph induced by views, triggers and functions}
\label{fig:dp}
\end{figure}
A database schema describes the structure of the data stored in a database system but also contains views, triggers, integrity constraints and user defined functions for data analysis.
Functions and these different rule types, namely deductive, active and normative rules, are typically defined with various interdependencies.
For example, views are defined with respect to base relations and/or some other views inducing a hierarchy of derived queries.
In particular, the expression {\tt  CREATE VIEW q AS SELECT \ldots \ FROM p$_1$,p$_2$,...,p$_n$} leads to the set \{p$_1$ $\rightarrow$ q, \ldots, p$_n$ $\rightarrow$ q\} of direct dependencies where q is a  derived relation and p$_i$ denote either a derived or a base relation.
These direct dependencies are typically represented by means of a predicate dependency graph which allows for analyzing indirect dependencies, too. Those indirect dependencies allow for understanding the consequences of changes made to the instances of the given database schema (referred to as update propagation in the literature) or to its structure.
Understanding the consequences of structural changes of a base table, for example, is important if a database user wants to know all view definitions potentially affected by these changes. 

Various dependencies are provided by the rules and functions in a database schema such as table-to-table dependencies induced by triggers or views-to-table dependencies which can be induced by functions.
A sample dependency graph is given in Figure~\ref{fig:dp} depicting dependencies between the base relations \{b$_1$, b$_2$, b$_3$\}, the derived relations \{v$_1$, v$_2$, v$_3$\}, the triggers \{t$_1$, t$_2$\}, and the functions \{f$_1$, f$_2$\}. 
For example, trigger t$_2$ fires upon changes in b$_2$ and refers in its action part to b$_3$ whereas
function f$_2$ is called from v$_3$ and executes operations affecting b$_3$.
The transitive closure allows for detecting indirect dependencies such as the one between b$_2$ and b$_3$ due to the path b$_2$ $\rightarrow$ v$_2 \rightarrow$ v$_3$ $\rightarrow$ f$_2$ $\rightarrow$ b$_3$ . 

\begin{figure}[t]
\center
\begin{tikzpicture}
\draw (-4.6,1) rectangle (1.14,-5.4);
\node[text width=8.8cm] () at (0,0) {\begin{tt}CREATE VIEW v$_3$ AS \newline
(SELECT name, AVG(sum), f$_1$() \newline
\hspace*{1pt} FROM v$_2$\newline
\hspace*{1pt} GROUP BY name); \end{tt}};
\node[text width=6.8cm] () at (0,-1.3 ) {\ldots};
\node[text width=8.8cm] () at (0,-3.5) {\begin{tt}CREATE FUNCTION f$_1$() \newline
 RETURN datetime IS \newline
 PRAGMA AUTONOMOUS\_TRANSACTION; \newline
 BEGIN \newline
 \hspace*{5pt}time := NOW(); \newline
 \hspace*{5pt}INSERT INTO b$_3$ (time\_log)\newline
 \hspace*{15pt}VALUES (time); \newline
 \hspace*{5pt}RETURN time; \newline
 END; \end{tt}};

\node[text width=8.8cm] () at (6.9,-0.7) {\begin{tt}v$_3$(X,Y,Z) \m{\lif} v$_2$(X,W), Z=f$_1$(), \newline
\hspace*{55pt} Y=f('avg',W,X). \vspace*{7pt} \newline 
derived(v$_3$, 3). \newline
dep(v$_3$, v$_2$). \newline
attr(v$_3$, 1, 'name'). \newline
call(v$_3$, \{1\}, 2, 'avg'). \newline
call(v$_3$, $\emptyset$, 3, f$_1$). \end{tt}};

\node[text width=6.8cm] () at (6.9,-2.8) {\ldots};

\node[text width=8.8cm] () at (6.9,-4) {\begin{tt}func(f$_1$, 0). \newline
dep(b$_3$, f$_1$). \newline
ftype(f$_1$, 'auto\_trans').
 \end{tt}};

\end{tikzpicture}
\caption{Translating an SQL view and function into a Datalog Meta-Program}
\label{fig:translation}
\end{figure}

This analysis can be further refined by structural details (e.g., negative vs. positive dependencies as needed in approaches for update propagation) as well as by considering the syntactical components of 
schema objects such as column names (attributes) or operator types (sum, avg, insert, delete, etc.).  
To this end, the definitions of schema objects need to be parsed and the obtained tokens stored as queryable facts. 
This kind of analysis is well-known from meta-programming in Prolog which led to the famous vanilla interpreter~\cite{HiLl89}.
For readability reasons we use Datalog instead of Prolog or SQL. A sample translation is given in Figure~\ref{fig:translation} where Datalog facts of the form 
\begin{verbatim}
dep(To,From)  % dependency relation (path/2 its transitive closure)    
base(R,A)     % base relation R with arity A
derived(V,A)  % view V with arity A
call(V,I,O,F) % input I and output O of function F in view V
attr(R,P,N)   % position P of attribute named N in relation R
\end{verbatim}
are used (amongst others) for representing meta-information about the given view and user-defined function.
Based on these facts, the analysis of schema elements can be simply realized by means of Datalog queries like
\begin{tabbing}
{\tt attr\_dups(R1,R2,N) $\leftarrow$ attr(R1,\_,N),attr(R2,\_,N),R1<>R2.}\\
{\tt idb\_func\_pred(V) $\leftarrow$ derived(V,\_),call(V,\_,\_,\_).}\\
{\tt base\_changes(B) $\leftarrow$ path(B,f$_1$),base(B,\_),func(f$_1$,\_).}\\
{\tt tbl\_dep(A,B) $\leftarrow$ base(A,\_),base(B,\_),path(A,F),path(F,B),func(F,\_).}
\end{tabbing}
for determining reused attribute names, views calling a function, base tables possibly changed by function f$_1$, and cyclic dependencies between two base tables through a function.
In doing so, many interesting properties of schema elements can be systematically determined which supports users in understanding the interrelationships of schema elements. 
Most database systems already allow for storing and querying meta-data about schema elements in a simple way but a comprehensive (and in particular user-driven) analysis is missing. In contrast, the proposed Datalog program can be easily extended by user-defined rules and the respective program can be directly transfered into a given database using SQL. This way, the schema analysis becomes a natural part of a database application.

\section{Functional Dependency Propagation}
\label{Functional Dependency Propagation}

In order to show the expressiveness of our approach, we investigate the possibility to compute induced FDs for derived relations using the deductive rules introduced above. 
FDs form special constraints which are assumed to hold for any possible valid database instance.
The FD propagation problem analyzes how FDs are propagated through the dependency graph of a database.
The problem is undecidable in the general setting for arbitrary relational expressions~\cite{Klug80} and even restricted to SC views, i.e., relational expressions allowing selection and cross product only, it turns out to be coNP-complete.\footnote{An in-depth discussion on the complexity can be found in~\cite{Fan08}.}
In favor of addressing the general setting,  we drop the ambition of achieving completeness by considering a special case, only. 
Instead, we allow for arbitrary expressions over all relational operators, multiple propagation steps and possibly finite domains\footnote{Finite domains cause troubles as FDs can occur due to the fact of limited possible value combinations.} in order to cover the majority of practical cases.
Due to the inherent complexity of the necessary reasoning process, the FD propagation problem is well-suited for showing the expressiveness of using deductive rules for schema analysis.

\subsection{Preliminaries}
\label{Preliminaries}
%
A functional dependency $\alpha$ \m{\lthen} $B$ states that the attribute values of~$\alpha$ determine those of~$B$.
The restriction to univariate right sides can be done without loss of generality as multivariate right sides can be composed using Armstrong's composition axiom~\cite{Arm74}.
We allow $\alpha = \emptyset$ which means that the attribute values of~$B$ are constant and restrict the represented FDs to those satisfying $B \notin \alpha$ (omitted FDs can be retrieved via Armstrong's augmentation axiom).

For our FD propagation rules, we employ a Datalog variant with special data types for finite, one-leveled sets (with the corresponding set operations union~$\cup$, intersection~$\cap$, set \mbox{minus~$\setminus$}, as well as the check for empty sets $\alpha \neq \emptyset$) and finite, possibly nested lists (with comparison and manipulation functionality).
We use lower case letters for predicate names and constants, capital letters for variable names and Greek letters for sets. For example, the Datalog expression
\begin{center}
\begin{tt} 
p(L[1,2],$\varepsilon$) \m{\lif} b(L[1,2],$\alpha$,Y), c(Y,$\gamma$,Z), $\varepsilon$ = $\alpha \cup (\gamma \setminus \{$Z$\})$.
\end{tt}
\end{center}
defines a join between \begin{tt}b\end{tt} and \begin{tt}c\end{tt} where the first attribute of \begin{tt}p\end{tt} is a list comprising two elements 1 and 2, and the second attribute of \begin{tt}p\end{tt} is the union of the sets $\alpha$ and $\gamma$ without the value in \begin{tt}Z\end{tt}.

In our approach we use the extended transitivity axiom 
\begin{equation} \alpha \lthen B, \ \gamma \lthen D, \ B \in \gamma, \ D \notin \alpha \ \Rightarrow \ \alpha \cup (\gamma \setminus B) \lthen D \label{eq:extended transitivity} \end{equation}
to derive transitive FDs.
Note that if $B \notin \alpha$ and $D \notin \gamma$, then the derived FD also satisfies $D \notin \alpha \cup (\gamma \setminus B)$.

\subsubsection{Rule Normalization}
\label{Rule Normalization}

In order to provide a systematic approach for FD propagation, we need the input rules to be in a so-called normal form, where each rule corresponds to exactly one relational operator. Any set of Datalog rules can be transformed into an equivalent set of normalized rules while properties of the original rule set like being stratifiable are preserved~\cite{BM08}:
\begin{example} \label{ex:rule normalization}
The rule \begin{tt}p(W,Z) \m{\lif} s(W,X),$\;$t(X,Y,Z),$\;$Y=2.\end{tt} can be normalized as
\begin{center}
\begin{tabular}{lcl}
\begin{tt} p(W,Z)     \end{tt} &\m{\lif}& \begin{tt} q(W,X,Y,Z).\end{tt} \\
\begin{tt} q(W,X,Y,Z)\end{tt} &\m{\lif}& \begin{tt} r(W,X,Y,Z),$\;$Y=2.\end{tt} \\
\begin{tt} r(W,X,Y,Z)\end{tt} &\m{\lif}& \begin{tt} s(W,X),$\;$t(X,Y,Z).\end{tt} \\
\end{tabular}
\end{center}
where the first rule corresponds to the projection operator~$\pi$, the second rule is a selection~$\sigma$ with constraint \begin{tt}Y=2\end{tt}, and the third rule represents the join~$\Join$ of \begin{tt}s\end{tt} and \begin{tt}t\end{tt}. 
\end{example}

In the following, we assume that the Datalog rules defining views are transformed into normal form for further analysis.
That is, each Datalog rule corresponds to exactly one of the following relational algebra operators (\{X$_h$\}, \{Y$_i$\}, \{Z$_j$\} denote pairwise disjoint variables):\\[-20pt]

\subsubsection{Projection $\pi$:}
\begin{tt}p(X$_1$,\small \ldots \normalsize,X$_k$) \m{\lif} q(Y$_1$,\small \ldots \normalsize,Y$_n$).\end{tt} \\ \hspace*{2.54cm}for \{X$_1$,\ldots,X$_k$\} $\subseteq$ \{Y$_1$,\ldots,Y$_n$\}\\[-25pt]

\subsubsection{Extension $\pi'$:}
\begin{tt}p(X$_1$,\small \ldots \normalsize,X$_k$,Y$_1$,\ldots,Y$_n$) \m{\lif} q(X$_1$,\small \ldots \normalsize,X$_k$).\end{tt} with equality \hangindent=2.46cm conditions \begin{tt}Y$_i$ = X$_j$\end{tt} or \begin{tt}Y$_i$ = 'a$_i$'\end{tt} for $1 \leq i \leq n$ and constants \begin{tt}a$_i$\end{tt}\\[-25pt]

\subsubsection{Selection $\sigma$:}
\begin{tt}p(X$_1$,\small \ldots \normalsize,X$_k$) \m{\lif} q(X$_1$,\small \ldots \normalsize,X$_k$), <Condition(X$_1$,\small \ldots \normalsize,X$_k$)>.\end{tt}\\[-25pt]

\subsubsection{Cross product $\times$:}
\begin{tt}p(X$_1$,\small \ldots \normalsize,X$_k$) \m{\lif} q(Y$_1$,\ldots,Y$_n$), r(Z$_1$,\small \ldots \normalsize,Z$_m$).\end{tt}\\ \noindent
\hspace*{3.13cm} for \{X$_1$,\ldots,X$_k$\} $=$ \{Y$_1$,\ldots,Y$_n$\} $\dot{\cup}$ \{Z$_1$,\ldots,Z$_m$\}\\[-25pt]

\subsubsection{Union $\cup$:}
\begin{tt}p(X$_1$,\small \ldots \normalsize,X$_k$) \m{\lif} p$_i$(X$_1$,\small \ldots \normalsize,X$_k$).\end{tt} for $1 \leq i \leq 2$\\[-25pt]

\subsubsection{Intersection $\cap$:}
\begin{tt}p(X$_1$,\small \ldots \normalsize,X$_k$) \m{\lif} q(X$_1$,\small \ldots \normalsize,X$_k$), r(X$_1$,\small \ldots \normalsize,X$_k$).\end{tt}\\[-25pt]

\subsubsection{Negation $-$:}
\begin{tt}p(X$_1$,\small \ldots \normalsize,X$_k$) \m{\lif} q(X$_1$,\small \ldots \normalsize,X$_k$), not r(X$_1$,\small \ldots \normalsize,X$_k$).\end{tt}\\[-25pt]

\subsubsection{Join} $\Join$\textbf{:}
\begin{tt}p(X$_1$,\small \ldots \normalsize,X$_k$) \m{\lif} q(Y$_1$,\small \ldots \normalsize,Y$_n$), r(Z$_1$,\small \ldots \normalsize,Z$_m$).\end{tt}\\ \noindent
\hspace*{1.3cm} for \{X$_1$,\ldots,X$_k$\} $=$ \{Y$_1$,\ldots,Y$_n$\} $\cup$ \{Z$_1$,\ldots,Z$_m$\}\\[-5pt]

\noindent In order to simplify the FD propagation, we will not allow for self joins or cross products (i.e., \mbox{\begin{tt}q$\;=\;$r\end{tt}}) which can always be achieved by applying renaming of one of the respective relations first.

\subsection{Representation of FDs and Normalized Rules}
\label{Representation of FDs}

We assume that functional dependencies for EDB predicates are given in a relation \begin{tt}edb\_fd(p,\m{\alpha},B,ID)\end{tt}.
Here \m{\alpha} and B are (sets of) column numbers of the relation~\begin{tt}p\end{tt}.
The fact represents the functional dependency $\alpha$ \m{\lthen} $B$ for the relation~\begin{tt}p\end{tt}.
The \begin{tt}ID\end{tt} is of type list and used to identify the dependency in later steps, e.g., in case of union.
The derived functional dependencies will be represented in the same way in an IDB predicate~\begin{tt}fd(p,\m{\alpha},B,ID')\end{tt}.
Here \begin{tt}ID'\end{tt} is related to the dependency's ID where the FD is derived from for propagated FDs or to a newly created ID for FDs that arise during the propagation process.

As in normal form every rule corresponds to exactly one operator, we can refine the above defined dependency relation \begin{tt}dep/2\end{tt} to \begin{tt}rel/3\end{tt} by adding the respective operator.
A fact \begin{tt}rel(p,q,op)\end{tt} indicates that a relation~\begin{tt}p\end{tt} depends (positively) on a relation~\begin{tt}q\end{tt} via a relational operator~\begin{tt}op\end{tt} which is one of \begin{tt}'projection'\end{tt}, \begin{tt}'extension'\end{tt}, \begin{tt}'selection'\end{tt} \begin{tt}'product'\end{tt}, \begin{tt}'join'\end{tt}, \begin{tt}'negation'\end{tt},  \begin{tt}'intersection'\end{tt}, and \begin{tt}'union'\end{tt}.

We further introduce an EDB predicate \begin{tt}pos(head,body,pos\_head,pos\_body)\end{tt} for storing information on how the positions of non position preserving operators (cf.~Table~\ref{FDproperties}) transform from rule body to head (since FDs are represented via column numbers).
For the rules of Example~\ref{ex:rule normalization} these are:\\[-5pt]

\begin{tabular}{lcl}
\begin{tt}p(W,Z) \m{\lif} q(W,X,Y,Z) \end{tt} & \m{\lthen} & \begin{tt} pos(p,q,1,1). pos(p,q,2,4). \end{tt} \\[5pt]
 \begin{tt}r(W,X,Y,Z) \m{\lif} s(W,X),t(X,Y,Z) \end{tt} & \m{\lthen} & \begin{tt} pos(r,s,1,1). pos(r,s,2,2).\end{tt} \\

&  & \begin{tt} pos(r,t,2,1). pos(r,t,3,2). \end{tt} \\
&  & \begin{tt} pos(r,t,4,3). \end{tt} \\[3pt]
\end{tabular}\\
Remembering that each relation is defined via one operator only and that we exclude self joins for simplicity (cf.\ Section~\ref{Rule Normalization}), the above defined relation \begin{tt}pos/4\end{tt} is non-ambiguous.
Finally, we have two EDB predicates \begin{tt}eq(pred,pos1,pos2)\end{tt} and \begin{tt}const(pred,pos,val)\end{tt} for information on equality conditions (e.g., \begin{tt}X\,=\,Y\end{tt} or \begin{tt}X\,=\,const\end{tt} resp.) in extension and selection rules.




%
%
%
%
%

\subsection{Propagation Rules}
\label{Propagation Rules}

In this section, we present three different types of propagation rules for (a)~propagating FDs to the next step, (b)~introducing additional FDs arising from equality constraints, and (c)~calculating transitive FDs. 

\begin{example}
\label{ex:FD propagation}
Consider again the rule set introduced in Example~\ref{ex:rule normalization}. If we assume two FDs \begin{tt}fd(s,\{1\},2,ID$_1$)\end{tt} and \begin{tt}fd(t,\{1,2\},3,ID$_2$)\end{tt} for the base relations \begin{tt}s\end{tt}~and~\begin{tt}t\end{tt} we obtain the following propagation process (omitting IDs).
First, both FDs are propagated to \begin{tt}r\end{tt} resulting in \begin{tt}fd(r,\{1\},2,-)\end{tt} and \begin{tt}fd(r,\{2,3\},4,-)\end{tt} (with the appropriate column renaming for the latter FD). By transitivity we have \begin{tt}fd(r,\{1,3\},4,-)\end{tt} as a combination of the two.
All three FDs are propagated to~\begin{tt}q\end{tt} together with \begin{tt}fd(q,$\emptyset$,3,-)\end{tt} resulting from the equality constraint~ \begin{tt}Y=2\end{tt}.
Applying transitivity results in three more FDs for~\begin{tt}q\end{tt}, but only \begin{tt}fd(q,\{1\},4,-)\end{tt} is propagated further to~\begin{tt}p\end{tt} as \begin{tt}fd(p,\{1\},2,-)\end{tt}.
The complete list of propagated FDs including IDs is given in Example~\ref{ex:with IDs}.
\end{example}


Table~\ref{FDproperties} summarizes the properties of how FDs are propagated via the different relational operators which form the basis for the propagation rules.
(Note that in the first two rows attention has to be paid for \textit{no}/--, whereas in the last three rows special attention has to be paid for \textit{yes}/$\times$.)
In most cases, the FDs are propagated as they are (with adjustments on the positions for $\pi$, $\times$, and $\Join$).
If there is a single rule defining a derived relation, the source FDs transform to FDs for the derived relation (restricted to the attributes in use).
Union forms an exception where even common FDs are only propagated in special cases and is therefore treated separately (cf. Section~\ref{Union and Recursion}).
For extensions and selections where additional FDs can occur due to equality conditions as well as for joins transitive FDs may appear so that taking the transitive closure becomes necessary.
In cases where the number of tuples is reduced (i.e., $\sigma$, $\cap$, $\Join$, and $-$) it is possible that new FDs appear as there are less tuples for which the FD constraint must be satisfied.
But as we are working on schema and not on instance level, this FD had to be present in a specific part of the parent relations.
The only case, where a FD is propagated to just a part of the derived relation is union.\\

\renewcommand{\arraystretch}{1.5}
\begin{table}[t]
\begin{center}
\begin{tabular}{m{4.8cm}|c|c|c|c|c|c|c|c}
Properties                                           & \ \ $\pi$ \ \ & \ \ $\pi'$ \ \ & \ \ $\sigma$ \ \ & \ \ $\times$ \ \ & \ \ $\cup$ \ \ & \ \ $\cap$ \ \ & \ \ $-$ \ \ & \ \ $\Join$ \ \ \\
\hline \hline
FDs are preserved                               & \ \ \ $\times$*$_1$ & $\times$ & $\times$ & $\times$ & -- & $\times$ & \ \ \ $\times$*$_2$ & $\times$ \\
 \hline
Positions are preserved                        & --               & $\times$ & $\times$ &\ \ \ --*$_3$ & $\times$ & $\times$ & $\times$ & --  \\
 \hline
Transitive FDs can appear                    & --               & $\times$     & $\times$     & --               & --               & --               & --               & $\times$     \\
  \hline
Additional FDs from equality conditions (variables and constants)
                                                         & --               & $\times$     & $\times$     & --               & --               & --               & --               & --               \\
 \hline
Additional FDs caused by instance reduction may appear
                                                         & --               & --               & $\times$     & --               & --               & $\times$     & $\times$     & $\times$     \\
 \hline
\end{tabular}
\end{center}

$\times$ $\hat{=}$  yes, -- $\hat{=}$ no

*$_1$: those where all contained variables are maintained

*$_2$: those of the minuend

*$_3$: positions of the first factor are preserved, positions of the second factor get an offset\\[-5pt]
\caption{Properties of FD propagation categorized by operator}
\label{FDproperties}
\end{table}

\renewcommand{\arraystretch}{1}

The different propagation rules for all relational operators except union, additional FDs due to equality conditions, and transitive FDs are specified in the following.
The definition and usage of IDs and how they are propagated is deferred to Section~\ref{Union and Recursion}.

\subsubsection{(a) Induced FDs}

For direct propagation of FDs from one level to the next, we distinguish between \textit{position preserving} and \textit{non position preserving} operators.
In the first case FDs can be directly propagated as they are (\ref{eq:position preserving}), whereas in the latter adjustments on the column numbers are necessary (\ref{eq:non position preserving}). 
The two EDB predicates \begin{tt}pos\_pres\end{tt} and \begin{tt}non\_pos\_pres\end{tt} comprise the respective operators as listed in Figure~\ref{list_operators}.
\begin{equation} \label{eq:position preserving} \verb+fd(P,+\alpha\verb+,B,-)+ \lif \verb+fd(Q,+\alpha\verb+,B,-), rel(P,Q,op), pos_pres(op).+\end{equation}
\begin{equation}\verb+fd(P,{X+_1\verb+,+\ldots\verb+,X+_n\verb+},Y,-)+ \lif \verb+fd(Q,{A+_1\verb+,+\ldots\verb+,A+_n\verb+},B,-),+ \hspace*{1cm} \label{eq:non position preserving}
\end{equation}\\[-15pt]
\begin{tt}
 \hspace*{1.95cm} pos(P,Q,X$_1$,A$_1$),\ldots, pos(P,Q,X$_n$,A$_n$), pos(P,Q,Y,B),\\
 \hspace*{2cm} rel(P,Q,op), non\_pos\_pres(op).\end{tt}

\begin{figure}[t]
\centering
\begin{tabular}{l}
\begin{tt}pos\_pres('selection').\end{tt}\\
\begin{tt}pos\_pres('extension').\end{tt}\\
\begin{tt}pos\_pres('negation').\end{tt}\\
\begin{tt}pos\_pres('intersection').\end{tt}\\
\end{tabular}%
\hspace*{1cm}%
\begin{tabular}{l}
\begin{tt}non\_pos\_pres('projection').\end{tt}\\
\begin{tt}non\_pos\_pres('product').\end{tt}\\
\begin{tt}non\_pos\_pres('join').\end{tt}\\
\end{tabular}
\caption{Position preserving (left) and non position preserving (right) operators.}
\label{list_operators}
\end{figure}

\subsubsection{(b) Additional FDs}

For any equality constraint \begin{tt}X\,=\,Y\end{tt} we can deduce the dependencies $X \lthen Y$ and $Y \lthen X$ as after the application of the constraint the values of $X$ and $Y$ coincide. Similar, a constant constraint \begin{tt}X\,=\,const\end{tt} induces the dependency $\emptyset \lthen X$.
Translated to our approach that is for any fact \begin{tt}eq(R,pos1,pos2)\end{tt} and \begin{tt}const(R,pos,val)\end{tt} respectively we derive the following FDs:%
\begin{equation} \verb+fd(R,{pos1},pos2,-)+ \lif \verb+eq(R,pos1,pos2).+\label{eq:equality1} \end{equation}%
\begin{equation} \verb+fd(R,{pos2},pos1,-)+ \lif \verb+eq(R,pos1,pos2).+\label{eq:equality2} \end{equation}%
\begin{equation} \verb+fd(R,+\emptyset\verb+,pos,-)+ \lif \verb+const(R,pos,val).+\label{eq:constant} \end{equation}%

\subsubsection{(c) Transitive FDs}

Since transitive FDs can only arise for certain operators it is sufficient to deduce transitive FDs in those cases (cf.~Table~\ref{FDproperties}).
For the computation we use the following two rules:
\begin{equation} \verb+fd(P,+\varepsilon\verb+,D,-)+ \lif \verb+fd(P,+\alpha\verb+,B,-), fd(X,+\gamma\verb+,D,-),+ \hspace*{2.25cm} \label{eq:transitive}\end{equation}\\[-20pt]
\begin{tt} \hspace*{3.8cm} B $\in \gamma$, D $\notin \alpha$, $\varepsilon = \alpha \cup (\gamma \setminus\{$B$\})$, trans(P). \end{tt}
\begin{equation} \label{eq:chase} \verb+fd(P,{X},Y,-)+ \lif \verb+fd(P,+\alpha\verb+,X,ID), fd(P,+\alpha\verb+,Y,ID), trans(P).+\end{equation}
The first rule implements the extended transitivity axiom (\ref{eq:extended transitivity}) and the second equates the right sides of two identical FDs (identified by matching IDs) with the same left side.
The IDB predicate \begin{tt}trans/1\end{tt} comprises those relations where transitive FDs may occur. Base relations are included to start with a complete set of representatives.
\begin{center}
\begin{tabular}{lll}
\begin{tt}trans(R)\end{tt} & \m{\lif} & \begin{tt}base(R,\_).\end{tt}\\
\begin{tt}trans(R)\end{tt} & \m{\lif} & \begin{tt}rel(R,\_,'join').\end{tt}\\
\begin{tt}trans(R)\end{tt} & \m{\lif} & \begin{tt}eq(R,\_,\_).\end{tt}\\
\begin{tt}trans(R)\end{tt} & \m{\lif} & \begin{tt}const(R,\_,\_).\end{tt}\\
\end{tabular}
\end{center}





\subsection{Union and Recursion}
\label{Union and Recursion}

In case of union $p = p_1\,\cup\,p_2$ even common FDs of $p_1$ and $p_2$ are only propagated in special cases.
Consider the following example of student IDs. 
For each university, the student ID uniquely identify the student associated with it.
But the same student ID can be used by different universities for different students.
So although we have the FD \textit{student ID \m{\lthen} student name} in the relations \begin{tt}Bonn\_students\end{tt} and \begin{tt}Cologne\_students\end{tt}, it is not a valid FD in the union of both.
A common FD of $p_1$ and $p_2$ is only propagated to $p$ if the domains of the FD are disjoint, or if they match on common instances.
The first case can only be handled safely on schema level if constants are involved.
The latter is the case if the FDs have the same origin and are propagated in a similar way.
Whether two FDs have the same origin can be easily checked with the \begin{tt}path\end{tt} relation of Section~\ref{Schema Analysis}.
This criteria is not yet enough as the FDs might have been manipulated during the propagation process (e.g., changes in the ordering, equality constraints, etc.).
So we employ a system of identifiers to track those changes made to a certain FD.
For the IDs we use a list structure that adopts the tree structure of~\cite{Klug80} who represents FDs as trees with source domains as leaves and the target domain as the tree's root.
As the target is already handled in the FD itself, we keep track of the source domains and transitively composed FDs, only.

At the beginning, each FD $\alpha \lthen B$ gets a unique identifier ID$_i$.
The idea is to propagate this ID together with the FD and to keep track of the modifications made to that FD.
For this purpose we attach an ordered tuple, a (possibly nested) list to the ID, i.e., ID$_i[A_1,\ldots, A_n]$ for the above base FD with $\alpha = \{A_1,\ldots, A_n\}$.
For the position preserving operators (that in particular do not change the FD's structure) the ID is identically propagated in (\ref{eq:position preserving}).
For the non position preserving operators the positions are updated (using a UDF) similarly to the position adjustments of the FD itself in (\ref{eq:non position preserving}).
The difference is that the ID maintains an ordering and that the cardinality stays invariant.
For constant constraints, we set the constant value as ID in (\ref{eq:constant}), equality constraints in (\ref{eq:equality1}), (\ref{eq:equality2}) and (\ref{eq:chase}) get the (column number of the) left side as ID.
In (\ref{eq:transitive}) we replace the occurrences of the column number \begin{tt}B\end{tt} in the ID of  \begin{tt}fd(X,$\gamma$,D,-)\end{tt} by the ID of \begin{tt}fd(X,$\alpha$,B,-)\end{tt}.
Note that this corresponds to a column number replacement for equality constraints.

Our propagation rules including IDs are the following:
\begin{equation} \tag{2'} \verb+fd(P,+\alpha\verb+,B,ID)+ \lif \verb+fd(Q,+\alpha\verb+,B,ID), rel(P,Q,op), pos_pres(op).+\end{equation}\\[-24pt]
\begin{equation} \tag{3'} \verb+fd(P,{X+_1\verb+,+\ldots\verb+,X+_n\verb+},Y,ID<X+_i\verb+>)+ \lif \verb+fd(Q,{A+_1\verb+,+\ldots\verb+,A+_n\verb+},B,ID<A+_i\verb+>),+ \hspace*{1cm} 
\end{equation}\\[-15pt]
\begin{tt}
 \hspace*{1.95cm} pos(P,Q,X$_1$,A$_1$),\ldots, pos(P,Q,X$_n$,A$_n$), pos(P,Q,Y,B),\\
 \hspace*{2cm} rel(P,Q,op), non\_pos\_pres(op).\end{tt}
\begin{equation} \tag{4'} \verb+fd(R,{pos1},pos2,pos1)+ \lif \verb+eq(R,pos1,pos2).+ \end{equation}\\[-24pt]
\begin{equation} \tag{5'} \verb+fd(R,{pos2},pos1,pos2)+ \lif \verb+eq(R,pos1,pos2).+ \end{equation}\\[-20pt]
\begin{equation} \tag{6'} \verb+fd(R,+\emptyset\verb+,pos,val)+ \lif \verb+const(R,pos,val).+ \end{equation} 
\newpage
\begin{equation} \tag{7'} \verb+fd(P,+\varepsilon\verb+,D,ID+_\varepsilon\verb+)+ \lif \verb+fd(P,+\alpha\verb+,B,ID+_{\alpha}\verb+), fd(X,+\gamma\verb+,D,ID+_\gamma\verb+),+ \hspace*{1.35cm} \end{equation}\\[-18pt]
\begin{tt} \hspace*{3.8cm} B $\in \gamma$, D $\notin \alpha$, $\varepsilon = \alpha \cup (\gamma \setminus\{$B$\})$, trans(P) \end{tt} \\
\begin{tt} \hspace*{3.8cm} ID$_\varepsilon =$ replace(ID$_\gamma$,B,ID$_\alpha$).\end{tt}
\begin{equation} \tag{8'} \verb+fd(P,{X},Y,X)+ \lif \verb+fd(P,+\alpha\verb+,X,ID), fd(P,+\alpha\verb+,Y,ID), trans(P).+\end{equation}

\begin{example}
\label{ex:with IDs}
\vbox{
For the FD propagation in Example~\ref{ex:FD propagation} we have the following IDs:
\begin{center}
\begin{tabular}{m{0.9cm}cm{3.7cm}m{0.9cm}cm{3.4cm}}
\begin{tt}fd(s,\end{tt}&\begin{tt}\{1\},\end{tt}&\begin{tt}2,$\,$ID$_1$[1]).\end{tt}
& \begin{tt}fd(q,\end{tt}&\begin{tt}\{1\},\end{tt}&\begin{tt}2,$\,$ID$_1$[1]).\end{tt}\\
\begin{tt}fd(t,\end{tt}&\begin{tt}\{1,2\},\end{tt}&\begin{tt}3,$\,$ID$_2$[1,2]).\end{tt}
& \begin{tt}fd(q,\end{tt}&\begin{tt}\{2,3\},\end{tt}&\begin{tt}4,$\,$ID$_2$[2,3]).\end{tt}\\
&&& \begin{tt}fd(q,\end{tt}&\begin{tt}\{1,3\},\end{tt}&\begin{tt}4,$\,$ID$_2$[ID$_1$[1],3]).\end{tt}\\
\begin{tt}fd(r,\end{tt}&\begin{tt}\{1\},\end{tt}&\begin{tt}2,$\,$ID$_1$[1]).\end{tt}
& \begin{tt}fd(q,\end{tt}&\begin{tt}$\emptyset$,\end{tt}&\begin{tt}3,$\,$'3').\end{tt}\\
\begin{tt}fd(r,\end{tt}&\begin{tt}\{2,3\},\end{tt}&\begin{tt}4,$\,$ID$_2$[2,3]).\end{tt}
& \begin{tt}fd(q,\end{tt}&\begin{tt}\{2\},\end{tt}&\begin{tt}4,$\,$ID$_2$[2,'3']).\end{tt}\\
\begin{tt}fd(r,\end{tt}&\begin{tt}\{1,3\},\end{tt}&\begin{tt}4,$\,$ID$_2$[ID$_1$[1],3]).\end{tt}
& \begin{tt}fd(q,\end{tt}&\begin{tt}\{1\},\end{tt}&\begin{tt}4,$\,$ID$_2$[ID$_1$[1],'3']).\end{tt}\\ \\[-5pt]
&&& \begin{tt}fd(p,\end{tt}&\begin{tt}\{1\},\end{tt}&\begin{tt}2,$\,$ID$_2$[ID$_1$[1],'3']).\end{tt}
\end{tabular}\\[8pt]
\end{center}
}
\end{example}
%
%
A common ID implies that the same modifications (not counting column number shifts due to joins, projections and equality constraints) have been made to a common base FD.
This means that the FD is preserved in the case of union.
So the final propagation rule for union reduces to a check for similar IDs:
\begin{equation}
\verb+fd(P,+\alpha\verb+,B,ID)+ \lif \verb+fd(P+_1\verb+,+\alpha\verb+,B,ID), fd(P+_2\verb+,+\alpha\verb+,B,ID),+ \hspace*{2.5cm} \label{eq:union}
\end{equation}\\[-15pt]
\begin{tt}\hspace*{1.1cm}rel(P,P$_1$,'union'), rel(P,P$_2$,'union'), path(P$_1$,X), path(P$_2$,X).\end{tt}\\[5pt]
Common FDs resulting from equality and constant constraints are propagated in any case.
Due to the choice of IDs this case is captured in the above ID comparison rule.
Note that there are FDs with different IDs for with the common FD is maintained in the union. 
Examples of such cases which are not covered by our approach can be found in~\cite{Klug80}.

%
%
%
%
%
%
%

\subsubsection{Recursion}

Recursion forms another special case.
Consider the following recursive example modeling the ancestor relationship of a tree.
We have the following two rules recursively defining the relation \begin{tt}p\end{tt}:\\[-8pt]

\begin{tt}p(Child,$\,$Parent,$\,$Parent) \m{\lif} q(Child,$\,$Parent).\end{tt}

\begin{tt}p(Child,$\,$Parent,$\,$Ancestor) \m{\lif} p(Child,$\,$Parent,$\,$X),$\;$q(X,$\,$Ancestor).\end{tt}\\[5pt]
Or equivalently transformed into the normal form:
\begin{center}
\begin{tabular}{lcl}
\begin{tt}p(C,$\,$P,$\,$A)\end{tt}        & \ \ \m{\lif} \ \ & \begin{tt}p$_1$(C,$\,$P,$\,$A).\end{tt}\\
\begin{tt}p(C,$\,$P,$\,$A)\end{tt}        & \m{\lif} & \begin{tt}p$_2$(C,$\,$P,$\,$A).\end{tt}\\
\begin{tt}p$_1$(C,$\,$P,$\,$P)\end{tt}    & \m{\lif} & \begin{tt}q(C,$\,$P).\end{tt}\\
\begin{tt}p$_2$(C,$\,$P,$\,$A)\end{tt}    & \m{\lif} & \begin{tt}p$_3$(C,$\,$P,$\,$X,$\,$A).\end{tt}\\
\begin{tt}p$_3$(C,$\,$P,$\,$X,$\,$A)\end{tt} & \m{\lif} & \begin{tt}p(C,$\,$P,$\,$X), q(X,$\,$A).\end{tt}\\
\end{tabular}
\end{center}
Each vertex has a unique parent, so we assume the FD \textit{Child \m{\lthen} Parent} for~\begin{tt}q\end{tt}, i.e., \begin{tt}fd(q,\{1\},2,*[1])\end{tt}.
With the rules defined above we cannot compute any FD for \begin{tt}p\end{tt} although the FD \textit{Child \m{\lthen} Parent} is propagated from \begin{tt}q\end{tt}.
This is because the FD is propagated to the recursion's base case and survives the recursion step.
The rules above cannot detect this FDs as it is only propagated to \begin{tt}p$_1$\end{tt} but not derivable for \begin{tt}p$_2$\end{tt} or \begin{tt}p$_3$\end{tt}.

Our solution is to take the \textit{potential} FDs propagated to the recursion's base case and feed them into the recursion step(s).
If they \textit{survive} the recursion step, i.e., if they are propagated with the above defined rules (\ref{eq:position preserving})-(\ref{eq:union}) with identical ID, then they are propagated as FD for the whole relation.
Similar to the case of union it is important that the FD's ID is maintained, otherwise the recursion's union might destroy the FD property.

Since we limited the union rule in the definition of the normal form to two operands (cf.~Section~\ref{Rule Normalization}) the rule defining a linear recursive relation \begin{tt}P\end{tt} has only two components, which are w.l.o.g.\ a base case \begin{tt}Q\end{tt} and a recursive component \begin{tt}R\end{tt}.
We maintain this information in an EDB predicate \begin{tt}rec(P,Q,R)\end{tt}.
As potential FDs we introduce those that are propagated from the base case \begin{tt}Q\end{tt} to the base part of the recursive relation \begin{tt}P\end{tt}.
\begin{equation*}\verb+pfd(P,+\alpha\verb+,B,ID)+ \lif \verb+fd(Q,+\alpha\verb+,B,ID), rec(P,Q,R).+\end{equation*}
Finally after propagating them through the dependency graph, we can deduce those FDs present in the recursive relation by ID comparison: 
\begin{equation*}\verb+fd(P,+\alpha\verb+,B,ID)+ \lif \verb+fd(Q,+\alpha\verb+,B,ID), pfd(R,+\alpha\verb+,B,ID), rec(P,Q,R).+\end{equation*}
Applied to the ancestor example this results in the following facts:\\[-15pt]
\begin{center}
\begin{small}
\begin{tabular}{m{2.75cm}m{2.75cm}m{3.05cm}l}
\begin{tt}fd(q,1,2,*[1]).\end{tt}     & \begin{tt}pfd(p,1,2,*[1]).\end{tt} & \begin{tt}fd(p$_3$,3,4,*(3)).\end{tt}   & \begin{tt}pfd(p$_2$,1,2,*[1]).\end{tt}\\
                                   & \begin{tt}pfd(p,2,3,3).\end{tt} & \begin{tt}pfd(p$_3$,1,2,*[1]).\end{tt} & \begin{tt}pfd(p$_2$,2,3,*[2]).\end{tt}\\
\begin{tt}fd(p$_1$,1,2,*[1]).\end{tt} & \begin{tt}pfd(p,3,2,3).\end{tt} & \begin{tt}pfd(p$_3$,2,3,2).\end{tt} & \begin{tt}pfd(p$_2$,1,3,*[*[1]]).\end{tt}\\
\begin{tt}fd(p$_1$,2,3,2).\end{tt} & \begin{tt}pfd(p,1,3,*[1]).\end{tt}  & \begin{tt}pfd(p$_3$,3,2,3).\end{tt}\\
\begin{tt}fd(p$_1$,3,2,3).\end{tt} &                                 & \begin{tt}pfd(p$_3$,1,3,*[1]).\end{tt} & \begin{tt}fd(p,1,2,*[1]).\end{tt}\\
\begin{tt}fd(p$_1$,1,3,*[1]).\end{tt} &                                 & \begin{tt}pfd(p$_3$,2,4,*[2]).\end{tt} \\
                                   &                                 & \begin{tt}pfd(p$_3$,1,4,*[*[1]]).\end{tt}
\end{tabular}
\end{small}
\end{center}
The FD 1 \m{\lthen} 2, i.e., \textit{Child \m{\lthen} Parent}, is correctly derived for relation~\begin{tt}p\end{tt}.

\subsection{Discussion}
\label{Discussion}

In Section~\ref{Propagation Rules} we introduced our propagation rules for propagating functional dependencies.
To compute the set of propagated FDs these rules are simultaneously applied to the input Datalog program in normal form and the extracted meta-data tokens.
(Note that the relation \begin{tt}fd\end{tt} comprising the propagated functional dependencies is recursively defined.)
The rules are based on the observations in Table~\ref{FDproperties} which can be easily verified.
For example an equality constraint $X\!=\!Y$ introduces two FDs between the attributes $X$ and $Y$ as the attribute values of one determine the identical attribute values of the other.
Due to a new FD a transitive deduction of FDs inside the view may become possible, e.g., $X\!=\!Y,\ Y \lthen Z \Rightarrow X \lthen Z$.
If there is a functional dependency $\alpha \lthen B$ for which not all attributes of $\alpha \cup \{B\}$ are present in the projected relation~\begin{tt}p\end{tt}, then either one of the source domains or the target domain is missing, so the FD cannot be propagated to~\begin{tt}p\end{tt}.
If $A$ is missing in~\begin{tt}p\end{tt} and $\alpha \setminus \{A\} \lthen B$ is a valid FD in \begin{tt}p\end{tt}, then the FD has a matching FD in the parent relation and is propagated to~\begin{tt}p\end{tt} via this FD.


The propagated functional dependencies of our approach are not complete as the problem is undecidable in the general setting.
Also limited to a less expressive subset of the relational operators (e.g., restricted operator order SPC views) one has to assume the absence of finite domains to achieve completeness.
Nevertheless we are able to deal with many cases appearing in real world applications.

The FD propagation approach can be further generalized to allow user-defined functions \begin{tt}Y$_i$ = f(X$_{i_1}$,\ldots,X$_{i_n}$)\end{tt} in the extension operator $\pi'$.
As the EDB predicate \begin{tt}call/4\end{tt} already maintains the information which function \begin{tt}F\end{tt} is called in a view \begin{tt}V\end{tt} with input \begin{tt}I\end{tt} and output \begin{tt}O\end{tt}, we just can use the propagation rule
\begin{equation*}\verb+fd(V,I,O,F[I])+ \lif \verb+call(V,I,O,F).+\end{equation*}
to include this case as well.
After a function call, the output of the UDF is functionally determined by the input, i.e., \begin{tt}I \m{\lthen} O\end{tt}.

\subsection{Related Work}
\label{Related Work FD}

The work of~\cite{Klug80} was the first addressing the FD propagation problem.
In his paper, Klug considers relational algebra expressions and defines a set of rules which iteratively compute the set of induced FDs.
The rules are sound and --~limited to restricted operator order which is as powerful as relational algebra without set difference~-- complete, assuming the absence of finite domains.
The propagation with this approach is not complete anymore when considering more than one propagation step as in this case the restricted operator order might be violated which can cause the loss of FDs that reappear in later steps.
In any rule based approach, special attention has to be paid to the union operator.
In~\cite{Klug80} this difficulty is addressed with an algorithm systematically testing possible value combinations using formal values.

The authors of~\cite{Fan08} use the concept of \textit{conditional FDs} to handle union.
The idea is that the dependencies are not propagated to the whole view but are still maintained in its subsets.
In the student ID example above, the IDs are still unique restricted to one university.
So on the condition that the university is University of Bonn (Cologne), the FD \textit{student ID \m{\lthen} student name} can be propagated.

Besides the mentioned rule based approaches, \textit{the chase} is an established algorithm for FD implication.
Originally developed as lossless join test~\cite{ABU79}, the chase has been used to infer dependencies inside one relation~\cite{MMS79} and for FD-FD implication~\cite{BGHP98}.
Related to the approach of~\cite{Klug80} the idea is to equate formal values following and applying functional dependencies.
In~\cite{PBPP03} the chase is used to deal with FD propagation in linear recursive Datalog programs.

\section{Conclusion}
\label{Conclusion}

In this paper, we have discussed a uniform framework for analyzing schema objects using deductive rules. 
In doing so, many interesting schema object properties as well as relationships can be systematically deduced.
These form valuable information for database users.
Our approach can be easily extended for refining the analysis by
user-defined queries and is suitable for SQL systems due to the choice of Datalog.
The expressiveness of the underlying reasoning process has been demonstrated by the analysis of induced functional dependencies. 
The respective FD propagation problem is known to be undecidable and intricate in detail. 
And yet, the proposed deductive rules allow for covering known approaches to solving the FD propagation problem proposed so far. 
This includes multiple propagation steps, union and linear recursion showing the depth of analysis which can be achieved for schema analysis in this way.
Identifying how other FD propagation approaches can be integrated into our rule based approach as well as an experimental evaluation of our proposal is part of further research.

%
%
%


%



\end{document}